\documentclass[intlimits,twoside,a4paper]{article}

\usepackage{graphicx,amssymb,amsfonts,amsmath}

\usepackage[T2A]{fontenc}
\usepackage[cp1251]{inputenc}

\usepackage{cmpj2}

\issue{2012}{15}{4}{43004}

\doinumber{10.5488/CMP.15.43004}

\title[Entropy production in open quantum systems: exactly solvable qubit models]
{Entropy production in open quantum systems: exactly solvable qubit models}

\author[V.G. Morozov, G. R\"opke]{V.G. Morozov\refaddr{label1},
        G. R\"opke\refaddr{label2}}
\addresses{
 \addr{label1} Moscow State Technical  University of Radioengineering,
  Electronics, and Automation, \\
 Vernadsky Prospect 78, 119454 Moscow, Russia
 \addr{label2} University of Rostock, FB Physik,
   Universit\"atsplatz 3, D--18051 Rostock, Germany
 }

\authorcopyright{V.G. Morozov, G. R\"opke, 2012}
\date{Received July 3, 2012, in final form August 23, 2012}

\begin{document}

\maketitle

\begin{abstract}
We present analytical results for the time-dependent information entropy in
exactly solvable two-state (qubit) models. The first model describes
dephasing (decoherence) in a qubit coupled to a bath of harmonic
oscillators. The entropy production  for this model in the regimes of
 ``complete'' and ``incomplete'' decoherence is discussed.
 As another example,  we consider the damped Jaynes-Cummings model
 describing a spontaneous decay of a two-level system into the field
 vacuum. It is shown that, for all strengths of coupling,
 the open system passes through the mixed state with the maximum
 information entropy.
\pacs 03.65.Ud, 03.65.Yz
\end{abstract}

\section{Introduction}
The notion of information entropy has
 attracted a renewed interest over the last few decades in
connection with fundamental problems in the
theory of open quantum systems  and quantum entanglement
  (see, e.g.,~\cite{BreuerPetr02,Zurek03,Amico08,Horodecky09}
   and references therein).
Although the literature concerning diverse aspects of
 this topic is now quite voluminous,
 not much is known about the entropy behavior
  in concrete open systems exhibiting especially
   intriguing features of quantum dynamics: memory effects,
   dephasing (decoherence), entanglement, etc.

The simplest models describing  many fundamental dynamic properties of
open quantum systems are two-state systems. Such systems themselves
deserve thorough studies as the elementary carriers of quantum information
  (qubits)~\cite{Bouwmeester00,Valiev05}.
It is also important to note that some of two-state models admit
\textit{exact\/} solutions. The latter fact allows one to gain
a valuable insight into general properties of open quantum systems.
 From this point of view, it is of interest to analyze the time behavior of
  information entropy in exactly solvable models.

In this paper we present exact analytic results for
 the time-dependent information entropy in two physically reasonable
 qubit models which are frequently used in discussing different
 problems in the theory of open quantum systems.

\section{Entropy of a qubit}
 We consider a two-state quantum system (qubit) coupled to a
 reservoir. In what follows, the qubit and the reservoir will
 be referred to as subsystems A and B, respectively.

 Suppose that at time $t$ the state of the combined system
 (qubit plus reservoir)
  is described by some density matrix $\varrho^{}_{AB}(t)$. Then, the reduced
  density matrix of the qubit is defined as
  \begin{equation}
  \varrho^{}_{A}(t)= \text{Tr}^{}_{B}\varrho^{}_{AB}(t),
   \label{rho-A}
  \end{equation}
where $\text{Tr}^{}_{B}$ denotes the trace over the reservoir degrees of
freedom. The von Neumann (information) entropy of the qubit is given by
 \begin{equation}
  S^{}_{A}(t)= - \text{Tr}^{}_{A}\left[
 \varrho^{}_{A}(t) \ln \varrho^{}_{A}(t)
   \right].
  \label{S-A}
 \end{equation}

It is convenient to use the ``spin'' representation for a qubit by writing
 its orthonormal basis  $|0\rangle$ and $|1\rangle$ as
  \begin{equation}
  |0\rangle = \begin{pmatrix}
              0\\
              1
            \end{pmatrix},
            \qquad
 |1\rangle= \begin{pmatrix}
              1\\
              0
            \end{pmatrix}.
  \label{spin-rep}
  \end{equation}
Then, all operators referring to a qubit
 can be expressed in terms of the Pauli
  matrices $\vec{\sigma}=\left\{\sigma^{}_{1},\sigma^{}_{2},\sigma^{}_{3}\right\}$.
    In particular, the density matrix (\ref{rho-A}) can be written in
     the form~\cite{Allen75,Cohen77}
  \begin{equation}
  \varrho^{}_{A}(t)= \frac{1}{2}\left[
 1+ \vec{\sigma}\cdot \vec{v}(t)
   \right],
  \label{rho-v}
  \end{equation}
where
  \begin{equation}
 \vec{v}(t)= \text{Tr}^{}_{A}\left[\vec{\sigma}\,\varrho^{}_{A}(t)\right]
   \label{v(t)}
  \end{equation}
is the so-called Bloch vector. In~\cite{MorozovRoep12} we have shown that
there exists another representation for the qubit density matrix, which is
better suited to calculate
 $\ln\varrho^{}_{A}(t)$ in (\ref{S-A}):
     \begin{equation}
 \varrho^{}_{A}(t)= \frac{1}{2}\sqrt{1-v^{2}(t)}\,
 \exp\left[
    \vec{\sigma}\cdot \vec{u}(t)
     \right],
     \qquad
 u=\frac{1}{2}\,\ln\left(\frac{1+v}{1-v}\right).
    \label{rho-A-exp}
  \end{equation}
Here, $v(t)$ is the modulus of the Bloch vector and $\vec{u}\upuparrows
\vec{v}$. Strictly speaking, the representation (\ref{rho-A-exp}) is valid
only for a mixed state of a qubit with $v<1$. Note, however, that the
limit
 $v\to 1$ can be taken directly
  in the entropy~(\ref{S-A}) after calculating
 the trace. Using expressions (\ref{rho-A-exp}), one easily derives from
 (\ref{S-A})
  \begin{equation}
 S^{}_{A}(t)= \ln 2
  - \frac{1}{2}\left(1+v\right)\ln\left(1+v\right)
  - \frac{1}{2}\left(1-v\right)\ln\left(1-v\right).
    \label{S-qubit}
  \end{equation}
For a pure state ($v\to 1$), we have $S^{}_{A}=0$, as it should be. The
entropy has its maximum $S^{}_{A}=\ln 2$ in the mixed state
 with $v=0$, when the density matrix (\ref{rho-v}) is diagonal and
 $\langle 0|\varrho^{}_{A}|0\rangle=
 \langle 1|\varrho^{}_{A}|1\rangle=1/2$.

The square modulus of the Bloch vector (\ref{v(t)}) can in general be
written as
 \begin{equation}
 v^{2}(t)= 4 v^{}_{+}(t)v^{}_{-}(t) +v^{2}_{3}(t),
   \label{v2}
 \end{equation}
where
  \begin{equation}
  v^{}_{\pm}(t)= \text{Tr}^{}_{A}[\sigma^{}_{\pm}\varrho^{}_{A}(t)],
  \label{v+-def}
  \end{equation}
and $\sigma^{}_{\pm}=\left(\sigma^{}_{1}\pm i \sigma^{}_{2}\right)/2$. The
two terms in (\ref{v2}) have different physical interpretations.
 The quantities $v^{}_{+}(t)= \langle 0|\varrho^{}_{A}(t)|1\rangle$
  and $v^{}_{-}(t)=\langle 1|\varrho^{}_{A}(t)|0\rangle$ are
   often referred to as the \textit{coherences\/}. They describe the
 environmentally induced dephasing~\cite{BreuerPetr02}.
 On the other hand, the time behavior of the
component of the Bloch vector $v^{}_{3}=
  \langle 1|\varrho^{}_{A}(t)|1\rangle
   - \langle 0|\varrho^{}_{A}(t)|0\rangle $
is determined by energy exchange between an open system and its
environment, which is responsible for complete statistical equilibrium in
the combined system. Thus, formulas~(\ref{S-qubit}) and~(\ref{v2}) are
convenient for studying  the role of different relaxation mechanisms in
  the entropy production.

\section{Entropy production in a dephasing model}
We start with a simple spin-boson model describing a qubit
 coupled to a reservoir of harmonic
 oscillators~\cite{BreuerPetr02,Luczka90,Unruh95,Palma96}. The
  Hamiltonian of the model is (in our units $\hbar=1$)
  \begin{equation}
    H= H^{}_{A}+ H^{}_{B} + H^{}_{I}
    =\frac{\omega^{}_{0}}{2}\, \sigma^{}_{3}
    + \sum_{k}\omega^{}_{k}b^{\dagger}_{k}b^{}_{k}
   + \sigma^{}_{3}\sum_{k}
   \left(
        g^{}_{k} b^{\dagger}_{k} + g^{*}_{k}b^{}_{k}
   \right),
   \label{H-tot}
 \end{equation}
where $\omega^{}_{0}$ is the energy difference between the excited state
 $|1\rangle$ and the ground state $|0\rangle$ of the qubit.
   Bosonic  operators $b^{\dagger}_{k}$ and
  $b^{}_{k}$ correspond to the $k$th reservoir
   mode with frequency $\omega^{}_{k}$.

Note that $\sigma^{}_{3}$ commutes with the Hamiltonian (\ref{H-tot}). As
a consequence, the populations $\langle 0|\varrho^{}_{A}(t)|0\rangle$ and
 $\langle 1|\varrho^{}_{A}(t)|1\rangle$ do not depend on time.
 In other words, there is no relaxation to a complete equilibrium
 between the qubit and the environment; that is, the model is
 \textit{nonergodic\/}.
However, we shall see below that this model exhibits a dephasing
relaxation and entropy production  without energy exchange between the
qubit and the environment.

Let us first specify the initial density matrix
 $\varrho^{}_{AB}(0)$ of the combined system.
 Usually
 (see, e.g.,~\cite{BreuerPetr02,Luczka90,Unruh95,Palma96}) it is assumed
  that the subsystems $A$ and $B$ are uncorrelated,
  and the reservoir is in thermal equilibrium at some temperature $T$.
  In this paper we will concentrate on the entropy production
   in time-dependent \textit{entangled\/} quantum states of the combined
    system.
We assume that at time
 $t=0$, the combined system is prepared in a
   \textit{pure\/} quantum state   which is a direct product
    \begin{equation}
   |\psi^{}_{AB}(0)\rangle=
   \big(a^{}_{0}|0\rangle + a^{}_{1}|1\rangle\big)
   \otimes |0^{}_{B}\rangle,
      \label{PsiAB-0}
    \end{equation}
where $|a^{}_{0}|^2 + |a^{}_{1}|^{2}=1$, and
 $|0^{}_{B}\rangle$ denotes the ground state of the reservoir.
 The initial state $|0^{}_{B}\rangle$ is chosen only
   for simplicity's sake.
The subsequent discussion may easily be extended to the case of an
arbitrary initial state $|\psi^{}_{B}(0)\rangle$
 of the reservoir. The density matrix corresponding to
 (\ref{PsiAB-0}) is
  \begin{equation}
 \varrho^{}_{AB}(0)= |\psi^{}_{AB}(0)\rangle \langle\psi^{}_{AB}(0)|.
    \label{rhoAB-0}
  \end{equation}
Since the evolution of the combined system is unitary, the initial
 state (\ref{PsiAB-0}) evolves after time $t$ into the \textit{pure\/}
 state
  \begin{equation}
   |\psi^{}_{AB}(t)\rangle= \exp\left(-\ri Ht\right)\,|\psi^{}_{AB}(0)
   \rangle,
     \label{PsiAB-t}
  \end{equation}
so that the density matrix of the combined system is given by
  \begin{equation}
  \varrho^{}_{AB}(t)=|\psi^{}_{AB}(t)\rangle\langle\psi^{}_{AB}(t)|.
   \label{rhoAB-t}
  \end{equation}
In principle, the qubit density matrix $\varrho^{}_{A}(t)$, and then
 the entropy $S^{}_{A}(t)$ can be calculated by using (\ref{rhoAB-t}).
 It is more convenient, however, to calculate the modulus of
  the Bloch vector, $v(t)$, and then apply formula~(\ref{S-qubit}).
Since $v^{}_{3}$ is constant and is determined by the amplitudes
 $a^{}_{0}$ and $a^{}_{1}$ in (\ref{PsiAB-0}), we need only to
consider the coherences $v^{}_{\pm}(t)$. Note that expression
 (\ref{v+-def}) can be rewritten as
    \begin{equation}
   v^{}_{\pm}(t)= \text{Tr}^{}_{AB}\left[
    \sigma^{}_{\pm}(t)\varrho^{}_{AB}(0)
    \right] \, ,
    \label{v+-:Heis}
  \end{equation}
where $\sigma^{}_{\pm}(t)$ are the Heisenberg picture operators and the
trace is taken over all degrees of freedom of the combined system. In the
model (\ref{H-tot}), equations of motion for $\sigma^{}_{\pm}(t)$ can be
 solved exactly. The result reads~\cite{MorozovRoep12}
     \begin{equation}
    \sigma^{}_{\pm}(t)=
    \exp\left[   \pm \ri\omega^{}_{0}t \mp R(t)
        \right] \sigma^{}_{\pm}\, ,
   \label{sig-pm(t)}
  \end{equation}
where the operator $R(t)$  acts only on the reservoir states and is given
by
   \begin{equation}
 R(t)=
     \sum_{k}\left[
     \alpha^{}_{k}(t) b^{\dagger}_{k}-\alpha^{*}_{k}(t) b^{}_{k}
             \right],
             \qquad
  \alpha^{}_{k}(t)=
   2g^{}_{k}\frac{1-{\rm e}^{\ri\omega^{}_{k}t}}{\omega^{}_{k}}\, .
    \label{R(t)}
   \end{equation}
Substituting the expression (\ref{sig-pm(t)}) into (\ref{v+-:Heis}) and
 recalling formula (\ref{rhoAB-0}), we find
  \begin{equation}
 v^{}_{\pm}(t)=v^{}_{\pm}(0)\exp\left[\pm \ri\omega^{}_{0}t
  -\gamma^{}_{\text{vac}}(t)\right]
    \label{v+-(t)}
  \end{equation}
with the so-called \textit{vacuum decoherence
function\/}~\cite{BreuerPetr02}
 \begin{equation}
  \gamma^{}_{\text{vac}}(t)=
  -\ln \langle 0^{}_{B}|\exp[R(t)]|0^{}_{B}\rangle
  = - \sum_{k}\ln \langle 0^{}_{B}|
   \exp\big[\alpha^{}_{k}(t) b^{\dagger}_{k}-\alpha^{*}_{k}(t)
   b^{}_{k}\big]
   |0^{}_{B}\rangle.
   \label{gam-vac}
 \end{equation}
After simple algebra which we omit, we obtain
     \begin{equation}
 \gamma^{}_{\text{vac}}(t)=  \int^{\infty}_{0} \rd\omega\,
  J(\omega)\,
  \frac{1-\cos\omega t}{\omega^{2}}\, ,
      \label{gam-vac-contin}
   \end{equation}
 where the continuum limit of the
 reservoir modes is performed, and the spectral
 density $J(\omega)$ is introduced by the rule
   \begin{equation}
    \sum_{k} 4|g^{}_{k}|^{2}\,f(\omega^{}_{k})=
     \int^{\infty}_{0} \rd\omega\, J(\omega)f(\omega).
     \label{J(omega)}
   \end{equation}
Now using the solution~(\ref{v+-(t)}) and taking into account that
 $v(0)=1$, we find from equation~(\ref{v2})
  \begin{equation}
  v^{2}(t)=
   v^{2}_{3} + \left(1- v^{2}_{3}\right)\exp\left[-2\gamma^{}_{\text{vac}}(t)\right].
    \label{v-expl}
  \end{equation}
 Formulas (\ref{S-qubit}) and (\ref{v-expl}) determine the time
evolution of the qubit entropy. To go beyond these formal relations,
  one needs some information on the spectral density $J(\omega)$.
 In many cases of physical interest
  (see, e.g.,~\cite{Palma96,Legett87}),
 $J(\omega)$ may be considered to be
 a reasonably smooth function
 which has a power-law behavior  $J(\omega)\propto \omega^{s}$\ ($s>0$)
  at frequencies much less than some ``cutoff'' frequency
   $\Omega$, characteristic of the reservoir  modes. In the limit
    $\omega\to\infty$, $J(\omega)$ is assumed to fall off at least
    as some negative power of $\omega$.
     For the  spectral density, we shall take the expression
     which is most commonly used in the theory of spin-boson
 systems~\cite{Luczka90,Unruh95,Palma96,Legett87}:
    \begin{equation}
   J(\omega)= \lambda^{}_{s}\Omega^{1-s}\,\omega^{s}\,
    {\rm e}^{-\omega/\Omega}\, ,
    \label{J-model-exp}
  \end{equation}
where $\lambda^{}_{s}$ is a dimensionless coupling constant.
   The case $s=1$ is usually called the ``Ohmic'' case, the  case $s>1$ is
  ``super-Ohmic'', and the case $0<s<1$ is ``sub-Ohmic''.

 Substituting the spectral density (\ref{J-model-exp}) into
  (\ref{gam-vac-contin}) and doing standard integrals, one gets
   \begin{equation}
 \begin{array}{ll}
 \displaystyle
  \gamma^{}_{\text{vac}}(t)=\lambda^{}_{s} \Gamma(s-1)
  \left\{
  1- \frac{\cos\left[(s-1)\,\arctan(\Omega t)\right]}
          {\left(1+\Omega^{2}t^{2}\right)^{(s-1)/2}}\right\},
         &    \quad (s\not=1),\\[10pt]
         \displaystyle
   \gamma^{}_{\text{vac}}(t)=
  \frac{\lambda^{}_{1}}{2}\,\ln\left(1+\Omega^2 t^2\right),
  &
  \quad
  (s=1),
   \end{array}
   \label{Gamma-vac-s}
 \end{equation}
where $\Gamma(s)$ is the Euler gamma function. Some important properties
of $\gamma^{}_{\text{vac}}(t)$  can easily be seen
 directly from the above expressions. First, $\gamma^{}_{\text{vac}}(t)$
  is a monotonously increasing function of time for $s\leqslant 1$. Second,
  in the super-Ohmic case $\gamma^{}_{\text{vac}}(t)$
   has a long-time limit:
  \begin{equation}
  \gamma^{}_{\text{vac}}(\infty)\equiv
  \lim_{t\to\infty} \gamma^{}_{\text{vac}}(t)=
   \lambda^{}_{s}\Gamma(s-1),
   \qquad
    (s>1).
   \label{Gamma-vac-sup}
  \end{equation}
Finally, the $\gamma^{}_{\text{vac}}(t)$
  monotonously saturates to $\gamma^{}_{\text{vac}}(\infty)$ for
  $1<s\leqslant 2$ and is a nonmonotonous function of time for
   $s>2$.

\begin{figure*}[htb] 
 \begin{center}
 {\includegraphics[width=0.33\textwidth]{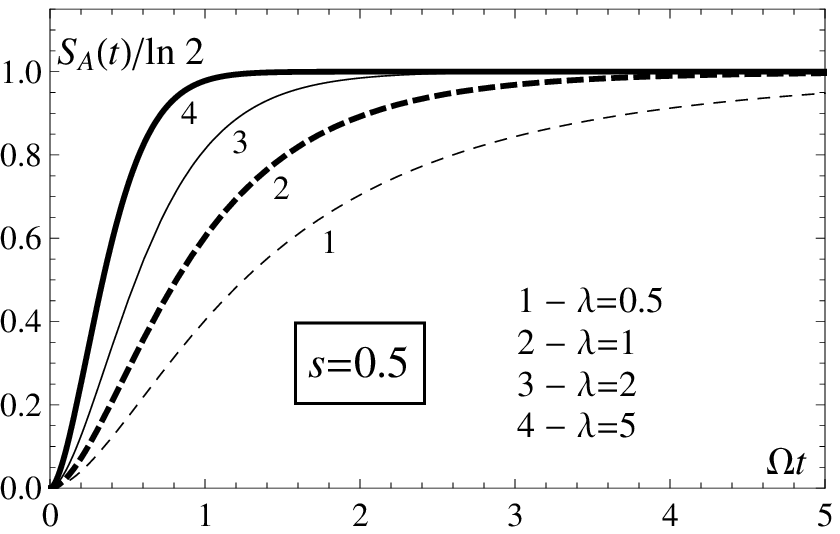}} 
 \hfill
 {\includegraphics[width=0.33\textwidth]{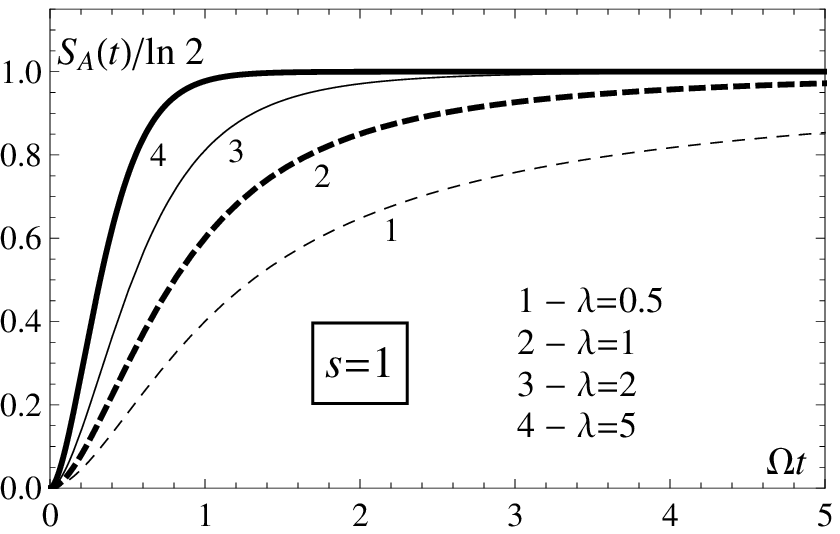}}
 \hfill 
 {\includegraphics[width=0.33\textwidth]{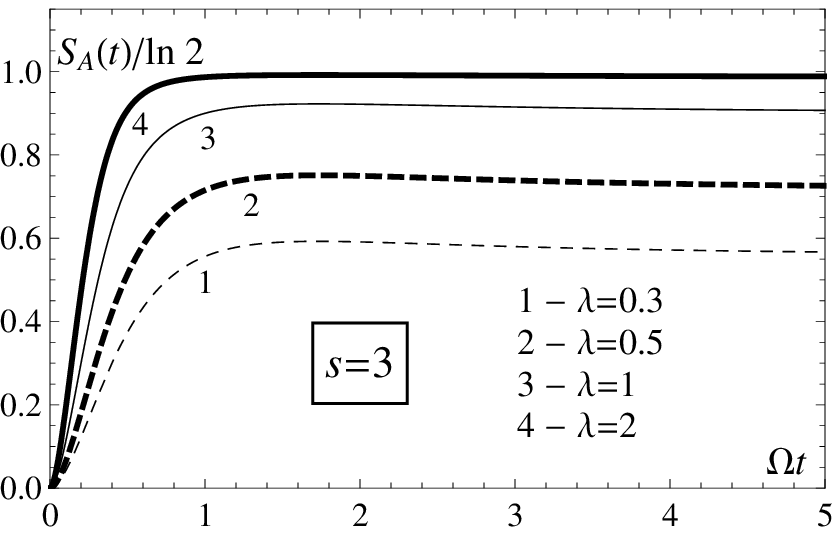}}
\end{center}
 \caption{Time evolution of the qubit entropy in the
 dephasing model (\ref{H-tot}). In all cases
 $\lambda\equiv\lambda^{}_{s}$.
 The qubit is initially
  in the pure state with equal populations ($v^{}_{3}=0$).}
 \label{Fig-Entropy-Dephas}
\end{figure*}

In discussing the properties of the qubit entropy in the model
(\ref{H-tot}),
  it is necessary to distinguish two cases:
 the regime of ``complete decoherence'', and the regime of
  ``incomplete decoherence''. In the former case ($s\leqslant 1$) we have
   $\gamma^{}_{\text{vac}}(t)\to \infty$ as $t\to\infty$, and hence
 $v^{}_{\pm}(t)\to 0$. Then, it follows directly from~(\ref{S-qubit}) and~(\ref{v-expl}) that the limiting value of
 the qubit entropy is given by
   \begin{equation}
  S^{}_{A}(\infty)=
   \ln 2
  - \frac{1}{2}\left(1+|v^{}_{3}|\right)
  \ln\!\left(1+|v^{}_{3}|\right)
  - \frac{1}{2}\left(1-|v^{}_{3}|\right)
  \ln\!\left(1-|v^{}_{3}|\right),
  \qquad
  (s\leqslant 1).
  \label{S-limit-compl}
   \end{equation}
The maximum qubit entropy $S^{}_{\text{max}}(\infty)=\ln 2$ corresponds to
 the initial state with equal populations
 ($v^{}_{3}=0$). In the case of ``incomplete decoherence'', we have
   \begin{equation}
    S^{}_{A}(\infty)=
   \ln 2 - \frac{1}{2}\left(1+v^{}_{\infty}\right)\ln\left(1+v^{}_{\infty}\right)
  - \frac{1}{2}\left(1-v^{}_{\infty}\right)\ln\left(1-v^{}_{\infty}\right),
  \qquad  (s>1),
    \label{S-limit-incompl}
   \end{equation}
where
  \begin{equation}
 v^{}_{\infty}=
   \left\{ v^{2}_{3}
   + \left(1 - v^{2}_3\right)
   \exp\left[-2\gamma^{}_{\text{vac}}(\infty)\right]\right\}^{1/2}.
   \label{v-infty}
  \end{equation}
Figure~\ref{Fig-Entropy-Dephas}\ illustrates
 the time behavior of the qubit
entropy for different values of the
 parameter $s$ and for different coupling strengths.
It should be emphasized at once that, for the case under consideration
here, the entropy production has no ``thermodynamic'' meaning. Indeed, at
any time $t$, the combined system is in the \textit{pure\/} quantum state
 (\ref{PsiAB-t}). Note, however, that this state is
  \textit{entangled\/}~\cite{BreuerPetr02}, i.e., it cannot be written as
  a direct product of states of the subsystems.
  Thus, the information entropy $S^{}_{A}(t)$ may be regarded as a measure of
   entanglement.

   Within the framework of the model (\ref{H-tot}) with the
    initial condition (\ref{PsiAB-0}), the structure of the
    entangled state (\ref{PsiAB-t}) can easily be established.
    Using the fact that the Hamiltonian (\ref{H-tot}) contains only
    the qubit operator $\sigma_{3}$, we obtain
     \begin{equation}
     |\psi^{}_{AB}(t)\rangle=
     a^{}_{0}{\rm e}^{-\ri\omega^{}_{0}t/2}
     |\psi^{(-)}_{B}(t)\rangle \otimes |0\rangle
     +a^{}_{1}{\rm e}^{\ri\omega^{}_{0}t/2}
     |\psi^{(+)}_{B}(t)\rangle \otimes |1\rangle
       \label{psiAB-dec}
     \end{equation}
with
    \begin{equation}
   |\psi^{(\pm)}_{B}(t)\rangle=
   \exp\big(-\ri H^{(\pm)}_{B}t\big)|0^{}_{B}\rangle,
   \qquad
   H^{(\pm)}_{B}=
   \sum_{k}\omega^{}_{k}b^{\dagger}_{k}b^{}_{k}
   \pm\sum_{k}
   \left(g^{}_{k} b^{\dagger}_{k} + g^{*}_{k}b^{}_{k}\right).
    \label{psiB+-}
    \end{equation}
The reservoir states $|\psi^{(\pm)}_{B}(t)\rangle$ are
 normalized but
 $\langle \psi^{(+)}_{B}(t) | \psi^{(-)}_{B}(t) \rangle\not=0$.
 It is possible, however, to write the states appearing in
 (\ref{psiAB-dec}) as linear combinations (the argument $t$ is
  omitted for brevity)
   \begin{equation}
  |\psi^{(\pm)}_{B}\rangle=
  \beta^{(\pm)}_{0}|\Phi^{}_{0}\rangle +
   \beta^{(\pm)}_{1}|\Phi^{}_{1}\rangle,
   \qquad
   |0\rangle=\alpha^{}_{0}|\phi^{}_{0}\rangle
   +\alpha^{}_{1}|\phi^{}_{1}\rangle,
   \qquad
  |1\rangle=\alpha^{\prime}_{0}|\phi^{}_{0}\rangle
   +\alpha^{\prime}_{1}|\phi^{}_{1}\rangle,
     \label{NewSt}
   \end{equation}
where the amplitudes may be determined from the following conditions:
 a) the pairs of states
  ($|\Phi^{}_{0}\rangle $,$|\Phi^{}_{1}\rangle $) and
  ($|\phi^{}_{0}\rangle $,$|\phi^{}_{1}\rangle$) are
   orthonormal; b) the terms with
   $|\Phi^{}_{0}\rangle\otimes|\phi^{}_{1}\rangle$ and
   $|\Phi^{}_{1}\rangle\otimes|\phi^{}_{0}\rangle$ cancel
 when expressions~(\ref{NewSt}) are inserted into~(\ref{psiAB-dec}). Then, the state of the combined system
   takes the form
   \begin{equation}
 |\psi^{}_{AB}(t)\rangle=
  A^{}_{0}(t)|\Phi^{}_{0}(t)\rangle\otimes|\phi^{}_{0}(t)\rangle
  +A^{}_{1}(t)|\Phi^{}_{1}(t)\rangle\otimes|\phi^{}_{1}(t)\rangle
     \label{psiAB-Schm}
   \end{equation}
with amplitudes satisfying
 $|A^{}_{0}(t)|^2+|A^{}_{1}(t)|^2=1$. Explicit expressions for
  $A^{}_{0}(t)$ and  $A^{}_{1}(t)$ are rather cumbersome and are not given here.

 Formula (\ref{psiAB-Schm})
  is an example of the so-called
  \textit{Schmidt decomposition\/} of quantum states of a combined
   system~\cite{BreuerPetr02}. Amplitudes $A^{}_{0}(t)$ and
   $A^{}_{1}(t)$ play a role of the corresponding
   \textit{Schmidt coefficients\/}. From (\ref{psiAB-Schm}) it follows
   that the reduced density matrices of the subsystems may be
    written as
     \begin{equation}
     \varrho^{}_{A}(t)=
      |A^{}_{0}|^2|\phi^{}_{0}\rangle\langle \phi^{}_{0}|
      + |A^{}_{1}|^2|\phi^{}_{1}\rangle\langle \phi^{}_{1}|,
      \qquad
    \varrho^{}_{B}(t)=
      |A^{}_{0}|^2|\Phi^{}_{0}\rangle\langle \Phi^{}_{0}|
      + |A^{}_{1}|^2|\Phi^{}_{1}\rangle\langle \Phi^{}_{1}|.
     \label{rhoArhoB}
     \end{equation}
Using these formulas, it is easy to show that
 \begin{equation}
 S^{}_{A}(t)=S^{}_{B}(t)=
  -|A^{}_{0}(t)|^2\ln |A^{}_{0}(t)|^2
  -|A^{}_{1}(t)|^2\ln |A^{}_{1}(t)|^2,
  \label{SA-SB-Schm}
 \end{equation}
 which is the well-known consequence
  of the Schmidt decomposition theorem for entangled quantum
   states~\cite{BreuerPetr02}.

\section{Spontaneous decay of a qubit}

Now we consider the entropy production in an exactly solvable  model which
describes a spontaneous decay of a two-level system
 into a field vacuum~\cite{BreuerPetr02,Garraway97,BreuerKappler99}. The total
 Hamiltonian of the model is
  \begin{equation}
 H=H^{}_{A}+ H^{}_{B} + H^{}_{I}=
 \frac{\omega^{}_{0}}{2}\, \sigma^{}_{3}
    + \sum_{k}\omega^{}_{k}b^{\dagger}_{k}b^{}_{k}
   + \sum_{k}\left(
        g^{}_{k} b^{\dagger}_{k}\sigma^{}_{-} + g^{*}_{k}b^{}_{k}\sigma^{}_{+}
   \right),
   \label{H-JCumm}
  \end{equation}
where the index $k$ labels the photon modes with
 frequencies $\omega^{}_{k}$.

The initial state of the combined system
 is again assumed to be given by (\ref{PsiAB-0}), i.e., $v(0)=1$ and,
 consequently, $S^{}_{A}(0)=0$.
 To obtain the entropy of the qubit at times $t>0$ we need to calculate
   $v(t)$. To do this, we will  closely follow the approach taken
    in the works~\cite{Garraway97,BreuerKappler99}.

It is convenient to work in the interaction picture
 with the unperturbed Hamiltonian $H^{}_{0}=H^{}_{A}+H^{}_{B}$.
Introducing  the interaction picture state vector
  $
  |\widetilde{\psi}^{}_{AB}(t)\rangle=
  \exp\left(\ri H^{}_{0}t\right)|\psi^{}_{AB}(t)\rangle,
  $
and applying the standard procedure, one obtains
   \begin{equation}
  |\widetilde{\psi}^{}_{AB}(t)\rangle=
  \exp^{}_{+}\left[-\ri\int^{t}_{0}\rd t'\, H^{}_{I}(t')\right]
   |\psi^{}_{AB}(0)\rangle,
     \label{psiAB-HI}
   \end{equation}
where $\exp^{}_{+}\left[ \cdots \right]$ is the chronologically ordered
 exponent, and
   \begin{equation}
 H^{}_{I}(t)= \sum_{k}
 \left(
  g^{}_{k}{\rm e}^{-\ri(\omega_0-\omega_k)t}b^{\dagger}_{k}\sigma^{}_{-}
  + g^{*}_{k}{\rm e}^{\ri(\omega_0-\omega_k)t}b^{}_{k}\sigma^{}_{+}
 \right)
    \label{HI(t)}
   \end{equation}
is the interaction picture Hamiltonian. As
 noted in the work~\cite{Garraway97}
 (see also~\cite{BreuerPetr02,BreuerKappler99}),
 there exists a simple representation for
  the state (\ref{psiAB-HI}), which can be derived
  using the following properties of the Hamiltonian~(\ref{HI(t)}):
   \begin{equation}
   \begin{array}{c}
  \displaystyle
    H^{}_{I}(t)\,|0\rangle\otimes|0^{}_{B}\rangle=0,
    \qquad
    H^{}_{I}(t)\,|1\rangle\otimes|0^{}_{B}\rangle=
    \sum_{k} g^{}_{k}{\rm e}^{-\ri(\omega_0-\omega_k)t}\,
   |0\rangle\otimes b^{\dagger}_{k}|0^{}_{B}\rangle,
    \\[7pt]
 \displaystyle
 H^{}_{I}(t)\,|0\rangle\otimes b^{\dagger}_{k}|0^{}_{B}\rangle=
 g^{*}_{k}{\rm e}^{\ri(\omega_0-\omega_k)t}\,
 |1\rangle\otimes|0^{}_{B}\rangle.
   \end{array}
     \label{HI-prop}
   \end{equation}
Recalling the expression (\ref{PsiAB-0}), one can easily show that at any time
$t$, the state (\ref{psiAB-HI}) is a superposition of
$|0\rangle\otimes|0^{}_{B}\rangle$,
 $|1\rangle\otimes|0^{}_{B}\rangle$, and
 $|0\rangle\otimes b^{\dagger}_{k}|0^{}_{B}\rangle$:
  \begin{equation}
 |\widetilde{\psi}^{}_{AB}(t)\rangle=
 \big[a^{}_{0}|0\rangle + c^{}_{1}(t)|1\rangle\big]
 \otimes|0^{}_{B}\rangle
  +\sum_{k}c^{}_{k}(t)|0\rangle\otimes
  b^{\dagger}_{k}|0^{}_{B}\rangle.
   \label{PsiAB-c}
  \end{equation}
The amplitudes $c^{}_{1}(t)$ and $c^{}_{k}(t)$ satisfy the initial
 conditions
 $c^{}_{1}(0)=a^{}_{1}$, $c^{}_{k}(0)=0$, and the normalization
  condition
   \begin{equation}
  |a^{}_{0}|^2 + |c^{}_{1}(t)|^2 + \sum^{}_{k}|c^{}_{k}(t)|^2=1.
    \label{Norm-cond}
   \end{equation}
The qubit density matrix in the interaction picture can now be calculated
 using (\ref{PsiAB-c}):
  \begin{equation}
   \widetilde{\varrho}^{}_{A}(t)=
   \text{Tr}^{}_{B}\left\{
  |\widetilde{\psi}^{}_{AB}(t)\rangle
  \langle \widetilde{\psi}^{}_{AB}(t)|
    \right\}=
     \frac{1}{2} +
   \frac{1}{2}\left(2|c^{}_{1}(t)|^2-1\right)\sigma^{}_{3}
   + a^{*}_{0}c^{}_{1}(t)\sigma^{}_{+}
   + a^{}_{0}c^{*}_{1}(t)\sigma^{}_{-} \, ,
 \label{rhoA-Int-sigma}
  \end{equation}
 where the amplitudes $c^{}_{k}(t)$ have been eliminated
 with the help of (\ref{Norm-cond}). Expressions for
  $v^{}_{\pm}(t)$ and $v^{}_{3}(t)$ follow immediately from (\ref{v(t)})
   and the relation
  $\varrho^{}_{A}(t)=
  \exp(-\ri H_{A}t)\widetilde{\varrho}^{}_{A}(t)\exp(\ri H_{A}t)$:
    \begin{equation}
   v^{}_{+}(t)={\rm e}^{\ri\omega^{}_0 t}a^{}_{0}c^{*}_{1}(t),
   \qquad
    v^{}_{-}(t)={\rm e}^{-\ri\omega^{}_0 t}a^{*}_{0}c^{}_{1}(t),
    \qquad
    v^{}_{3}(t)= 2|c^{}_{1}(t)|^2 -1.
     \label{v+-3}
    \end{equation}
Substituting these expressions into (\ref{v2}) and using
  $|a^{}_{0}|^2=1- |c^{}_{1}(0)|^2$, we finally obtain
   \begin{equation}
    v^{2}(t)= 1 - 4|c^{}_{1}(t)|^2
    \left[
  |c^{}_{1}(0)|^2 -|c^{}_{1}(t)|^2
    \right].
     \label{v-c1}
   \end{equation}
Thus, the modulus of the Bloch vector and, consequently, the qubit
 entropy $S^{}_{A}(t)$ are completely determined by
   $|c^{}_{1}(t)|^2$ which is the time-dependent population of the
   excited state $|1\rangle$.
It is important to note that the probability amplitude
 $c^{}_{1}(t)$ obeys a closed
 equation~\cite{BreuerPetr02,BreuerKappler99}
  \begin{equation}
  \dot{c}^{}_{1}(t)=
  - \int^{t}_{0} \rd t'\,f(t-t')c^{}_{1}(t')
  \label{c1-eq}
  \end{equation}
with the kernel
  \begin{equation}
  f(t)= \sum_{k}|g^{}_{k}|^2\,
  {\rm e}^{\ri(\omega^{}_{0}-\omega^{}_{k})t}
  \equiv
  \int \rd\omega\, J(\omega)\,
  {\rm e}^{\ri(\omega^{}_{0}-\omega)t},
   \label{c-f-eq}
  \end{equation}
where $J(\omega)$ is the spectral density of the field. In some important
special cases of $J(\omega)$, equation (\ref{c1-eq}) can be solved to give
 exact analytic solutions for
 $c^{}_{1}(t)$. We shall restrict ourselves to the so-called
  damped Jaynes-Cummings model
   (see, e.g.,~\cite{BreuerPetr02,BreuerKappler99})
  which describes the resonant coupling of a two-level atom to a single
  cavity mode of the field. In this model, the effective spectral density
  has the form
    \begin{equation}
  J(\omega)= \frac{1}{2\pi} \,
  \frac{\gamma^{}_{0}\lambda^2}
       {\left( \omega^{}_{0}-\omega\right)^2+\lambda^2},
    \label{Spectr-JCumm}
    \end{equation}
where $\lambda$ is a spectral width of the coupling, and
 the  parameter $\gamma^{}_{0}$
 defines the characteristic time scale $\tau^{}_{A}=1/\gamma^{}_{0}$
 on which the state of the qubit changes. For the details
 of solving the equation (\ref{c1-eq}) we refer
  to~\cite{BreuerPetr02,BreuerKappler99}. Here, we quote the resulting
  expression for the population $|c^{}_{1}(t)|^2$:
  \begin{equation}
 |c^{}_{1}(t)|^2=
  |c^{}_{1}(0)|^2 {\rm e}^{-\lambda t}
  \left[
   \cosh\left(\Lambda t/2\right)+\frac{\lambda}{\Lambda}\,
  \sinh\left(\Lambda t/2\right)
   \right]^2,
   \label{c2-JCum}
  \end{equation}
where $\Lambda=\left( \lambda^2 -2\gamma^{}_{0}\lambda\right)^{1/2}$.
Substituting the above expression into (\ref{v-c1}) and introducing
 the dimensionless coupling parameter
  \begin{equation}
 K=2\gamma^{}_{0}/\lambda,
   \label{JCum-K}
  \end{equation}
we obtain
  \begin{equation}
  v^2(t)= 1- 4|c^{}_{1}(0)|^4\,
   {\rm e}^{-\lambda t}F^{}_{K}(t)
   \left[
 1- {\rm e}^{-\lambda t}F^{}_{K}(t)
   \right].
    \label{v2-JCum-fin}
  \end{equation}
The form of the function $F^{}_{K}(t)$ depends on the value
 of $K$:
  \begin{equation}
   \begin{array}{ll}
   \displaystyle
   F^{}_{K}(t)=
   \left[\cosh\left(\sqrt{1-K}\,\lambda t/2\right)
   + \frac{1}{\sqrt{1-K}}\sinh\left(\sqrt{1-K}\,\lambda t/2\right)  \right]^2,
   & \qquad
   K<1,\\[10pt]
    \displaystyle
    F^{}_{K}(t)=
   \left[\cos\left(\sqrt{K-1}\,\lambda t/2\right)
   + \frac{1}{\sqrt{K-1}}\sin\left(\sqrt{K-1}\,\lambda t/2\right)  \right]^2,
   & \qquad
   K>1,\\[10pt]
   F^{}_{K}(t)=
   \left( 1+ \lambda t/2 \right)^2,
   & \qquad
   K=1.
   \end{array}
   \label{FK-JCum}
  \end{equation}
Formulas (\ref{S-qubit}) and (\ref{v2-JCum-fin})  completely determine the
qubit entropy in the Jaynes-Cummings model.
  \begin{figure*}[!h] 
 \begin{center}
 \text{a)}\hspace*{2pt}
{\includegraphics[width=6.2cm]{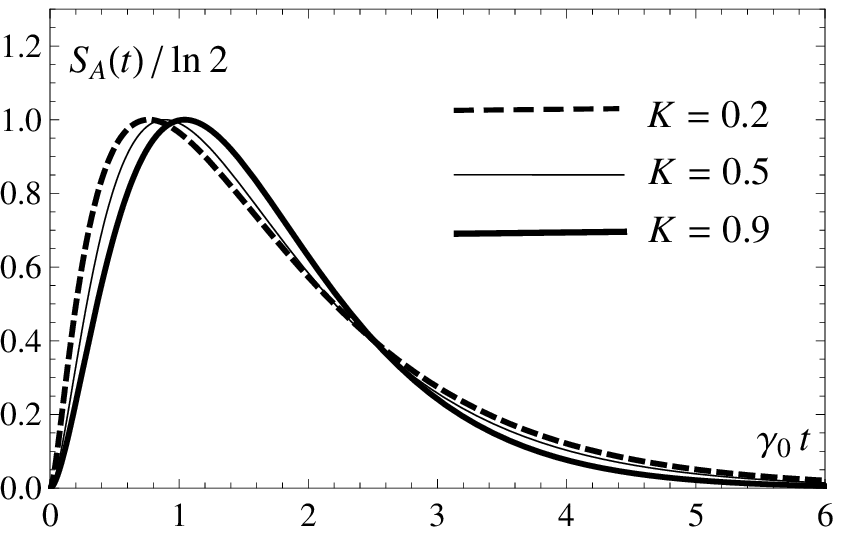}}
\hspace*{10pt}
\text{b)}\hspace*{2pt}
 {\includegraphics[width=6.2cm]{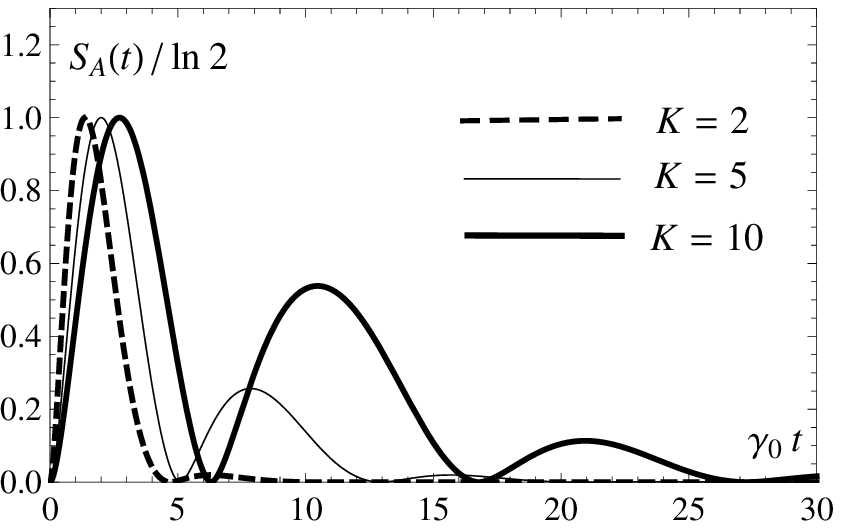}}
\end{center}
 \caption{Time evolution of the qubit entropy in the
 damped Jaynes-Cummings model: a) Weak and moderate coupling ($K<1$);
  b) Strong coupling ($K>1$). In all cases the qubit is initially
  in the excited state $|1\rangle$, i.e., $|c^{}_{1}(0)|^2=1$.}
 \label{Fig-Entropy-JCumm}
\end{figure*}
Figure~\ref{Fig-Entropy-JCumm} illustrates the time behavior of
$S^{}_{A}(t)$.
 We take the situation where the initial pure
 state $|1\rangle$ of the qubit evolves into the final
  pure state $|0\rangle$. Note that
   for all strengths of coupling, at some time
    $t^{}_\mathrm{m}$, the qubit  passes through  the mixed state
     with the \textit{maximum\/} entropy $S^{}_{A}(t^{}_\mathrm{m})=\ln 2$,
      i.e., with $v(t^{}_\mathrm{m})=0$.
   \begin{figure*}[!h] 
 \begin{center}
{\includegraphics[width=6.3cm]{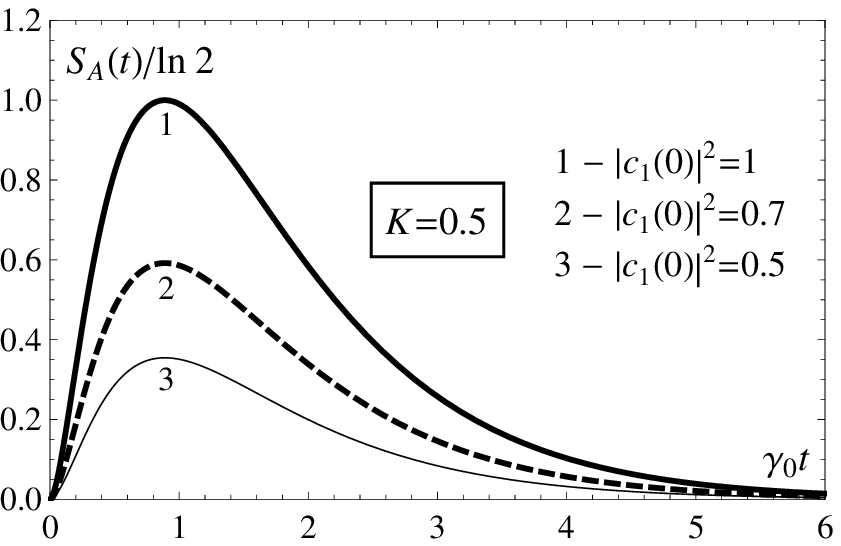}}
\hspace*{10pt}
 {\includegraphics[width=6.3cm]{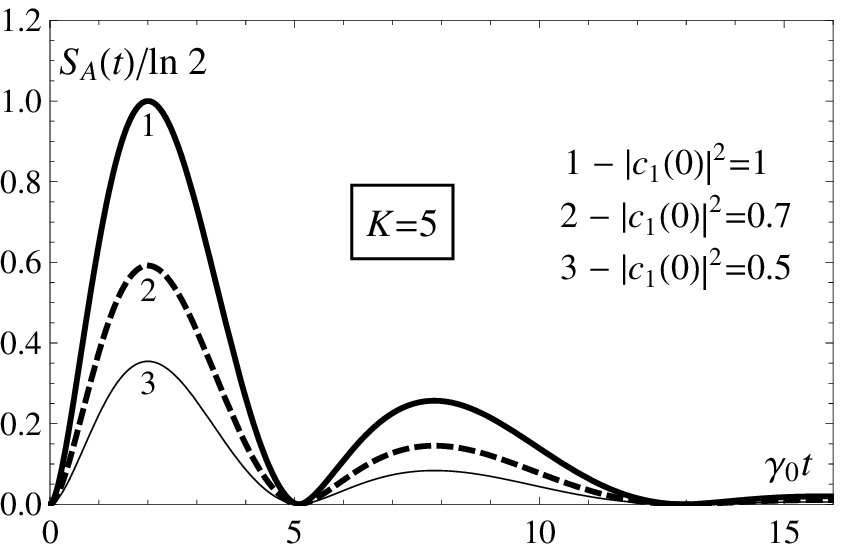}}
\end{center}
 \caption{Time evolution of the qubit entropy in the
 damped Jaynes-Cummings model for different initial
  populations of the excited qubit state  $|1\rangle$ in the cases
   of weak and strong coupling.}
 \label{Fig-Entropy-JCumm-c0}
\end{figure*}
It is interesting that this ``maximally entangled''
 intermediate state of the combined system
  appears  only in the case of the maximum
 initial population of the excited qubit state
 $|1\rangle$ ($|c^{}_{1}(0)|^2=1$),
as may be seen directly from~(\ref{v-c1}). The time behavior of the qubit
 entropy for smaller values of $|c^{}_{1}(0)|^2$ is shown in figure~\ref{Fig-Entropy-JCumm-c0}.

\section{Conclusion}
In this paper we have considered the time evolution of information
 entropy in two exactly solvable models of two-state open
  quantum systems (qubits). Calculations were based on
  the simple but quit general representation (\ref{S-qubit}) for the
   qubit entropy.

 Our discussion was restricted to the special case of
 the qubit and the reservoir being initially uncorrelated and the
  reservoir being in the ground state, i.e., at zero temperature.
  This simplifying assumption allows one to study in detail
  the entropy production in entangled time-dependent
  quantum states of a combined
   system. It should be noted, however, that
   in many situations of physical interest, a
   factorized (uncorrelated) initial state at $T=0$ cannot
    always be realized, so that the dynamics of
     thermal and correlated initial
    states, including the entropy behavior,
    is of great significance. For instance, it was shown
     in~\cite{MorozovRoep12} that the dephasing model
      (\ref{H-tot}) admits exact solutions for a large class
       of physically reasonable correlated initial states at finite
       temperatures. It was found that, for a sufficiently strong coupling,
     initial qubit-environment correlations have a profound effect
     on the dephasing process and on the time behavior of entropy.
 It would be of interest to study the
 entropy behavior in further examples of exactly
solvable models of open quantum systems in the presence of
 initial system-environment  correlations.

\section*{Acknowledgement}
This work  was supported by  DFG (Deutsche Forschungsgemeinschaft),
 SFB 652 (Sonderforschungsbereich~--- Collective Research Center 652).


\ukrainianpart

\title{Продукування ентропії у відкритих квантових системах: точно розв'язні кубіт моделі}

\author{В.Г. Морозов\refaddr{label1},
        Г. Репке\refaddr{label2}}
\addresses{
\addr{label1} Московський державний технічний університет  радіотехніки, електроніки та автоматики, \\
проспект Вернадського, 78, 119454 Москва, Росія
\addr{label2} Університет м. Росток, Університетська площа, 3, D--18051 м. Росток, Німеччина
 }

\makeukrtitle

\begin{abstract}
Ми представляємо аналітичні результати для часовозалежної
інформаційної ентропії в двостанових точно розв'язних (кубіт)
моделях. Перша модель описує дефазування (декогеренцію) в кубіті,
який є зв'язаний з резервуаром гармонічних осциляторів.
Обговорюється продукування ентропії для цієї моделі  у режимах
``повної'' та ``неповної'' декогеренції. Як інший приклад ми розглядаємо
задемпфовану модель Джейнса-Каммінгса, яка описує спонтанне
затухання дворівневої системи в польовому вакуумі. Показано, що для
всіх сил зв'язку, відкрита система переходить через змішаний стан з
максимумом інформаційної ентропії.
\keywords  інформаційна ентропія, відкриті квантові системи, кубіт
моделі, декогеренція, квантове заплутування
\end{abstract}

\end{document}